\documentclass{aastex631}
\usepackage[utf8]{inputenc}
\usepackage{textcomp}
\usepackage{gensymb}
\usepackage{amsmath}
\usepackage{amssymb}
\usepackage{xspace}
\usepackage{nicefrac}
\usepackage{xcolor}
\usepackage{graphicx}
\usepackage{soul}
\usepackage{hyperref}
\usepackage[caption=false]{subfig}
\usepackage{multirow}
\usepackage{cleveref}
\usepackage{natbib}
\usepackage[titletoc,toc,page]{appendix}
\usepackage{xcolor}
\usepackage{rotating}

\DeclareMathAlphabet\mathrsfso      {U}{rsfso}{m}{n}

\usepackage{esint}

\begin{document}

\title{On $\gamma$ rays as predictors of UHECR flux in AGNs}

\author{Cain\~a de Oliveira}
\email{caina.oliveira@usp.br}
\affiliation{Instituto de F\'isica de S\~ao Carlos, Universidade de S\~ao Paulo, Av. Trabalhador S\~ao-carlense 400, S\~ao Carlos, Brasil.}
\author{Rodrigo Guedes Lang}
\email{rodrigo.lang@fau.de}
\affiliation{Friedrich-Alexander-Universit\"at Erlangen-N\"urnberg, Erlangen Centre for Astroparticle Physics, Nikolaus-Fiebiger-Str. 2, 91058 Erlangen, Germany}
\author{Pedro Batista}
\email{pedro.batista@fau.de}
\affiliation{Friedrich-Alexander-Universit\"at Erlangen-N\"urnberg, Erlangen Centre for Astroparticle Physics, Nikolaus-Fiebiger-Str. 2, 91058 Erlangen, Germany}

\date{\today}

\begin{abstract}
    Active galactic nuclei (AGN) are among the main candidates for ultra-high-energy cosmic ray (UHECR) sources. However, while some theoretical and phenomenological works favor AGNs as the main sources, recent works have shown that using the very-high-energy (VHE) $\gamma$-ray flux as a proxy for the UHECR flux leads to a bad agreement with data. In this context, the energy spectrum and composition data are hardly fitted. At the same time, the arrival directions map is badly described and a spurious dipole direction is produced. In this work, we propose a possible solution to these contradictions. Using the observed $\gamma$-ray flux as a proxy may carry the implicit assumption of beamed UHECR emission and, consequently, its beam will remain collimated up to its detection on Earth. We show that assuming an isotropic UHECR emission and correcting the $\gamma$-ray emission proxy by Doppler boosting can overcome the problem. The combined fit of the spectrum and composition is improved, with a change of reduced $\chi^2$ from 4.6 to 3.1. In particular, the tension between the observed and modeled dipole directions can be reduced from $5.9 \ (2.1)\sigma$ away from the data to $3.5 \ (1.1)\sigma$ for $E > 8$ EeV ($E > 32$ EeV). We also show that this effect is particularly important when including AGNs of different classes in the same analysis, such as radio galaxies and blazars.
\end{abstract}

\section{Introduction}
The first detection of ultra-high energy cosmic rays (UHECRs)~\citep{PhysRevLett.10.146} raised significant questions about their origin. 
The discovery of astrophysical objects responsible for the acceleration of particles to ultra-high energies remains one of the most compelling mysteries in contemporary science~\citep{kotera2011astrophysics,alves-batista-2019}. 
The small flux of UHECRs requires experiments with a very large area, up to thousands of square kilometers, to improve the detection of events and minimize experimental uncertainties~\citep{RevModPhys.72.689}. 
The Pierre Auger Observatory~\citep{2015172} and Telescope Array~\citep{ABUZAYYAD201287} are the best examples of such feat with unprecedented exposure, leading to large statistics of high-quality data that allowed precision studies in the ultra-high-energy range.

UHECR arrival direction measurements have been extensively used in the search for UHECR accelerators. On the large-scale anisotropies, one of the most significant results comes in the form of the dipole measured by the Pierre Auger Observatory. The measured dipole reaches $6.8\sigma$ confidence level (CL) for events with energies exceeding 8 EeV and points outward the galactic center at ($\alpha = (97 \pm 8)^{\circ}$, $\delta = (-38^{+9}_{-9})^{\circ}$), which is powerful evidence for the dominance of extragalactic UHECR above this energy~\citep{Abdul_Halim_2024_19yr}.
For smaller scale anisotropies, \citet{Abreu_2022} reported a correlation between the arrival direction data of events with energies greater than $39$~EeV and a jetted active galactic nuclei (AGN) catalog with a confidence of $3.3\sigma$. The same analysis performed with a starburst galaxy (SBG) catalog reached $4.2\sigma$ confidence level. 
Combining Pierre Auger and Telescope Array data, a full-sky search for sources showed a correlation of 4.7$\sigma$ ($>38 / 49$~EeV for Auger/TA) with the SBG catalogue~\citep{Auger_TA_2023}.

To consider particle physics processes and cosmic magnetic field deflections involved during the travel from the source to Earth, Monte Carlo simulations have been developed~\citep{Batista_2016,Aloisio_2017}. 
The results of numerical propagation are compared with observations, and the free parameters of the model (nuclei fraction, spectral index, maximum energy, spectrum normalization) are constrained by fitting the simulation results to experimental data~(see, for example, \cite{Aab_2017,Eichmann_2022,Abdul_Halim_2024}). 

Although the acceleration mechanism is generally assumed to be identical for a given class of objects (i.e., assuming identical spectral index), different approaches have been used to predict the UHECR luminosity of each object~\citep{Aab_2018_indication,Eichmann_2018,Eichmann_2019,Eichmann_2022,Abreu_2022,deOliveira_2022,deOliveira_2023,Abdul_Halim_2024,Partenheimer_2024}.

Several works investigate AGNs as possible sources of UHECR. In radio galaxy scenarios, assuming the cosmic ray flux proportional to the jet power it is possible to phenomenologically explain some characteristics of the spectrum, composition, and arrival direction~\citep{Eichmann_2018,Eichmann_2019,Eichmann_2022,deOliveira_2022}. In particular, if Centaurus A, M87, and Fornax A are considered dominant sources, the high-energy dipole and small-scale anisotropies may be explained~\citep{deOliveira_2022,deOliveira_2023}. In the unified model for AGNs, radio galaxies are blazars whose jets are misaligned with our line of sight. If radio galaxies are sources of UHECR, it is natural to blazars be too. When the $\gamma$-ray luminosity ($L_{\gamma}$ ) is used as a proxy for the UHECR luminosity ($L_{CR}$) in an AGN catalog, the data is hardly fitted~\citep{Abdul_Halim_2024,Partenheimer_2024}. In this case, the main issue is that the strong signal from the jetted AGN Mkn 421 ($\sim130$~Mpc) generates a hotspot not observed in the data and dominates the dipole direction. In addition, the fit to the spectrum is worsened by the high contribution of distant sources.

In this work, we propose a possible way to conciliate these two views by reviewing the motivations behind using $L_\gamma$ as a proxy for $L_{\mathrm{CR}}$.
In section~\ref{sec:review} we present a comparison of the theoretical predictions for the acceleration of UHECRs in AGN jets and the origin of the $\gamma$-ray radiation. 
We demonstrate that $L_{\gamma}$ should be used carefully when applied as a weight to the UHECR flux. 
In section~\ref{sec:implications} it is shown that, when considering the intrinsic $\gamma$-ray luminosity rather than the observed $\gamma$-ray luminosity, the combined fit of the spectrum and composition data improves, while the agreement between the predicted arrival directions and data also enhances. 
The main results and outlook of this work are summarized in section~\ref{sec:conclusions}.

\section{Reviewing gamma rays and cosmic rays in jets} \label{sec:review}

The deflections suffered by cosmic rays during the trajectory from the accelerator to the detection prevent the direct identification of their sources. 
Without reliable models of the extragalactic and Galactic magnetic fields, theoretical arguments are combined with constraints of cosmic magnetic fields in the source search.
The usual approach for investigating possible individual sources focuses on the so-called local sources, with distances of about a hundred Mpc, for which the emitted UHECR will not have been as diffused as for farther sources and may still maintain some information about the source location~\citep{lang2021}. 
It is common to consider every source as a standard candle with an effective spectral index but with different cosmic ray luminosities, $L_{\rm{CR}}$. 
The observed luminosity in $\gamma$ rays, $L_{\gamma}^{\rm{obs}}$ has previously been used as a possible proxy~\citep{Aab_2018_indication,Abbasi_2018,Abreu_2022,PierreAuger:2023fcr,Abdul_Halim_2024,Partenheimer_2024}. 
However, a bad agreement between the model for AGNs and data is found with this assumption, in particular due to a strong contribution from Mkn~421, a blazar located at $\sim 130$ Mpc. 
In this section, we explore the assumption of using $L_{\gamma}^{\rm{obs}}$ as a proxy for $L_{\rm{CR}}$ in AGN and argue that the intrinsic luminosity $L_{\gamma}^{\rm{int}}$ may be a more robust assumption.

The \cite{hillas1984} condition is the minimum requirement when considering astrophysical objects as possible accelerators. 
The lobes, knots, and hotspots of jetted AGNs satisfy the Hillas condition for acceleration up to the ultra-high-energy scale~\citep{alves-batista-2019}. AGNs have been considered prime UHECR source candidates in different contexts~\citep{kimura_2018,matthews2018,Eichmann_2018,Eichmann_2019,Eichmann_2022,deOliveira_2022,rieger2022,deOliveira_2023}, in particular, due to the UHECR hotspot detected at $4\sigma$ by the Pierre Auger Observatory around the direction of Centaurus A~\citep{PierreAuger:2023fcr}.

Several sites in the structure of AGN have been proposed to be suitable for particle acceleration~(see \citet{MATTHEWS2020101543,rieger2022} for reviews), for example: the neighborhood of the supermassive black hole~\citep{Katsoulakos_2018,Coimbra-Araújo_2021,rita_BH_2022}; parsec- and kiloparsec-scale jet~\citep{kimura_2018,Rodrigues_2018,Seo_2023,Seo_2024}; backflow of the jet material~\citep{matthews2018}; the termination shock~\citep{Cerutti_2023}; and the lobes~\citep{osullivan2009}.

\subsection{Gamma rays in jets}
The broadband spectral energy distribution (SED) of jetted AGN has been measured from radio to $\gamma$ rays, with a characteristic double-peak shape. The detection of X-ray and $\gamma$ rays demonstrated the existence of regions of particle acceleration along AGN's jets~\citep[e.g.,][]{annurev:blandford}. The lower energy peak is normally attributed to synchrotron radiation from the interaction between accelerated electrons and positrons and magnetic fields of the medium~\citep[e.g.,][]{dermer_high_2009}.
As for the higher-energy peak, the most common hypotheses rely on the upscattering of low-energy photons to $\gamma$-ray energies, by higher-energy leptons, via inverse Compton (IC) processes.~\citep[e.g.,][]{Finke_2008}. 
These low-energy photons can originate from the synchrotron radiation emitted by the same particle population - known as synchrotron self-Compton (SSC) emission, or from thermal radiation coming from the broad and narrow line regions, the accretion disk, and the torus - known as external Compton (EC) emission~\citep[e.g.,][]{Böttcher_2013}.

The acceleration of hadrons also must occur at least as efficiently as electrons~\citep[e.g.,][]{ATOYAN2004381}. Radiation from hadronic origin, mainly from pion decays, may be added to the leptonic emission. Both proton-photon and proton-proton interactions can be responsible for $\gamma$ ray creation via neutral pion decay and successive electromagnetic cascades. In general, $\gamma$ rays from pion decay carry approximately $10\%$ of the energy of one high-energy proton. Due to the increased number of intermediate reactions, $\gamma$ rays originating from charged pion require higher proton energies. Then, the detection of $\gamma$ rays in the TeV scale can imply the existence of protons with energy in the PeV scale.~\citep{PhysRevD.74.034018,boettcher_relativistic_2012,Cao_2014}. 

It is important to note that for increasing $\gamma$-ray energies (starting at TeV) absorption via interactions with extragalactic background light (EBL) and cosmic microwave background (CMB) photons are more likely to occur, which greatly reduces the flux of $\gamma$~rays above such energies on Earth~\citep{de_angelis_2013,Gréaux_2024}.

The most common geometry of jet emission models consists of particles being accelerated in a compact region that is traveling at relativistic speeds down the jet.
In this region, also known as the \textit{blob}, the plasma of particles moves with a bulk Lorentz factor $\Gamma_b$ along the jet axis, while emitting photons isotropically in the \textit{blob}'s rest frame. For highly relativistic motions ($\Gamma_b \gg 1$), an isotropic emission in the \textit{blob}'s co-moving frame will be observed on Earth as a beamed emission, with a beaming angle $\theta_{\textrm{beam}} = \Gamma_b^{-1}$~\citep{dermer_high_2009}.

Relativistic transformations of the photon energy and emission angle will impact the observed photon flux. The intrinsic and observed energy of a photon, $\epsilon^{\rm{int}}_\gamma$ and $\epsilon^{\rm{obs}}_\gamma$ respectively, are related by the Doppler factor $\mathcal{D}$ defined as $\mathcal{D} \equiv \frac{\epsilon^{\rm{obs}}_\gamma}{\epsilon^{\rm{int}}_\gamma} = \big[ \Gamma_b\left( 1 - \beta_b \cos{\theta} \right) \big]^{-1}$,
where $\beta_b$ is the normalized velocity of the \textit{blob}, and $\theta$ is the angle between the jet axis and Earth's line of sight.

The observed photon flux is related to the intrinsic photon flux of a source, but is subject to relativistic transformations, being strongly dependent on the Doppler factor. The ratio between the observed flux density $F^{\rm{obs}}_\gamma$ of photons on Earth and the intrinsic photon flux density $F^{\rm{int}}_\gamma$ is 

\begin{equation}
\label{eq:spec_flux_ratio}    
    \frac{F^{\rm{obs}}_\gamma}{F^{\rm{int}}_\gamma} = \frac{\nu^{\rm{obs}}_\gamma}{\nu^{\rm{int}}_\gamma}\frac{d\nu^{\rm{int}}_\gamma}{d\nu^{\rm{obs}}_\gamma}\frac{dN^{\rm{obs}}_\gamma}{dN^{\rm{int}}_\gamma}\frac{dt^{\rm{int}}_\gamma}{dt^{\rm{obs}}_\gamma}\frac{d\Omega^{\rm{int}}_\gamma}{d\Omega^{\rm{int}}_\gamma},
\end{equation}
where $d\Omega = dA / d^2_L$, and $d_L^2$ is the invariant luminosity distance. From eq.~(\ref{eq:spec_flux_ratio}), it is possible to show that the observed flux of photons at Earth, $\Phi^{\rm{obs}}_\gamma$, and therefore the observed luminosity $L^{\rm{obs}}_\gamma$, is boosted by a factor $\mathcal{D}^4$ in relation to the intrinsic luminosity $L^{\rm{int}}_\gamma$, $L^{obs}_\gamma = \mathcal{D}^4 L^{int}_\gamma$.~\citep{boettcher_relativistic_2012}

\subsection{UHECR acceleration in AGNs}

UHECRs should be accelerated along the jet by different mechanisms in different regions~\citep{kimura_2018,MATTHEWS2020101543,rieger2022,Seo_2023,Seo_2024}. 
Magnetic reconnection can be present at the highly magnetized jet base~\citep{Medina-Torrejón_2021}. 
Diffusive shock acceleration should dominate in the shocked regions of the jet beam, backflow, and termination shock. 
Shear acceleration can occur in regions of high-velocity gradients, caused by the highly relativistic jet, and even in the neighborhood of the termination shock~\citep{Cerutti_2023}. 
Acceleration by second-order Fermi acceleration is possible in turbulent regions of the lobes. 
Only UHECR accelerated in the relativistic beamed plasma is subject to the beaming effect.

UHECR could be accelerated in shocks present in the jet.~\citep{Rodrigues_2018, Murase_Zhang_CenA_2023, zech_lemoine_2021}.
In general, highly relativistic shocks are not efficient UHECR accelerators, and mildly relativistic shocks are more promising~\citep{Lemoine_Pelletier_2010,10.1093/mnras/stu088, 10.1093/mnras/stx2485,matthews2018}. 
The cosmic ray emission must be isotropic in the shock rest frame, so relativistic beaming is expected in the laboratory rest frame. The shocks should accelerate not only cosmic rays but also electrons, which will radiate downstream of the shock, where the plasma Doppler factor will modulate the emission, as discussed in the previous section~\citep[e.g.,][]{zech_lemoine_2021}.
Even in mildly relativistic jets, the flux corrections due to Doppler factor can be significant, since the observed luminosity depends on $\mathcal{D}^4 = \big( \frac{1 + \beta}{1-\beta} \big)^2 \sim 3-80$, for $\beta \sim 0.5-0.8$ and line of sight jets. However, the angular distribution of UHECR is likely to be isotropized (in the lab frame) within the source region since these particles will cross the magnetized jet and the lobes before escaping, both with complex magnetic field structures, and the jet itself being subject to turbulences and the presence of knots~\citep{Goodger_2010,Dubey_2023,mignone_resistive_2023}.

Consider the propagation and possible acceleration of UHECR along the kpc-scale jet. 
Combining hydrodynamics and Monte Carlo simulations, \citet{Seo_2023,Seo_2024} found that the main mechanism accelerating UHECR above a few EeV is the relativistic shear acceleration at the interface jet-backflow. 
Combining magnetohydrodynamics with Particle-in-Cell simulations, \citet{Mbarek_2019} made a detailed study of the angular distribution of UHECR emission on the kpc-scale jet. The angular distribution of the accelerated UHECR depends mainly on the toroidal component of the jet magnetic field that can disperse (isotropic emission) or collimate (anisotropic emission) particles. 
The direction of emission will also be determined by the deflections inside the cocoon\footnote{Region of shocked material surrounding the jet~\citep[e.g.,][]{begelman1989overpressured}.}. In an anisotropic scenario, only about half of the particles were collimated inside an angle larger than $\Gamma_{jet}^{-1}$.

As AGN's jet inflates the lobes~\citep{agn_feedback,hardcastle_lobes_numerical,raise_code_jet_lobes}, the UHECR beam should cross them before reaching the extragalactic medium. Due to its extension and presence of turbulent/filamentary magnetic field \citep{carilli1996cygnus,Massaro_2011,hardcastle_lobes_numerical,meerkat_pictorA,ordered_B_radio_galaxies, fermi_lobes_cenA, wykes_2014_lobes, wykes_2015_lobes}, it is likely that the UHECR scatter inside the lobes, losing its directional information. The scattering length of a UHECR can be approximated as~\citep{osullivan2009,lang2020}
\begin{equation}
    \lambda_{scatt} \sim \kappa^2 \ell_c \Big( \frac{r_L}{\ell_c} \Big)^\delta,
\end{equation}
where $\kappa^2 = \nicefrac{B_0^2}{\delta B^2}$, $\ell_c$ is the coherence length of the magnetic field, $\delta$ is the diffusion coefficient, and
\begin{equation}
    r_L \approx \frac{E/\text{EeV}}{Z} \frac{1~\text{kpc}}{B_0/\mu \text{G}},
\end{equation}
is the gyroradius of the UHECR of charge $Z$. $B_0$ and $\delta B$ are the large-scale and turbulent components of the magnetic field respectively.

The lobes extend across $R \sim 100$~kpc with magnetic fields $\sim 1-10~\mu$G~\citep{Massaro_2011, fermi_lobes_cenA, wykes_2015_lobes,hardcastle_pictorA,meerkat_pictorA}. The coherence length of the magnetic field is assumed to be $\ell_c \sim 0.1 \times R \sim 10$~kpc~\citep{radioLobes_B20258,osullivan2009}. If $\lambda_{scatt}$ is smaller than $R$, the UHECR will suffer at least one scattering inside the lobes, losing its directional information. 
The major energy dissipation of the jet occurs in the $\sim$pc scale from the jet base~\citep{harvey2020powerful,shukla2020gamma}.
Then considering $\lambda_{scatt} \sim R$, we get the energy threshold for one scattering
\begin{equation}
    E_{scatt} \sim (Z \times 10~\text{EeV}) \ell_{10} B_{\mu G} \Big(\frac{10 R_{100}}{ \ell_{10} \kappa^2} \Big)^{1/\delta},
\end{equation}
where $\ell_{10} = \ell / 10~\text{kpc}$, $R_{100} = R / 100~\text{kpc}$, and $B_{\mu G} = B_0 / \mu\text{G}$. 
In a conservative estimation, we consider the high-energy non-resonant regime, $\delta = 2$~\citep{globus_2008}, and the fiducial value $\kappa \approx 1$ ($B_0 \sim \delta B$)~\citep{osullivan2009,wykes_2014_lobes}, 
\begin{equation}
    E_{scatt} \sim (Z \times 30~\text{EeV}) \delta B_{\mu G} \sqrt{\ell_{10} R_{100}}.
\end{equation}

Even if accelerated in relativistic blobs inside the jet, protons with energies up to $\sim 30$~EeV will lose their directional information in the source, by being isotropized while traveling through the lobes. 
However, due to limitations in the acceleration capacity of the sources, UHECR are unlikely protons up to such high energies. Assuming an electromagnetic origin for the UHECR acceleration, it is useful to introduce the magnetic rigidity $R = E/Z$. The combined fit performed by the Pierre Auger Collaboration~\citep{Abdul_Halim_2024_magfit} indicates a magnetic rigidity cutoff $R_{cut} \lesssim 10$~EV at the sources, and the flux at Earth being dominated by He and N above the ankle (5~EeV), with successive heavier composition. Taking $Z \sim 5$ we get $E_{scatt} \sim 150$~EeV, encompassing most of the detected UHECR. This value can still be larger since photodisintegration makes the composition lighter after propagation from the source.


This estimation depends on the position of the accelerator in the jet. The scattering can be inefficient if the acceleration occurs mainly on the termination shocks, observed as the hotspot found in FRII radio galaxies~\citep{Hardcastle_2007,Snios_2020}. 
As relativistic shocks, the termination shocks are poor accelerators of UHECR~\citep{10.1093/mnras/stx2552}. 
Yet, recently \citet{Cerutti_2023} found that particles can be efficiently accelerated up to $\sim10^{20}$~eV by crossing a cavity behind the termination shock. 
In this case it is unclear if a possible beamed emission will be sustained after leaving the acceleration region. The magnetic field of the vortex downstream of the cavity or of the hotspot itself might decollimate the beam, at least partially.

\section{Implications for the search for sources} \label{sec:implications}

If particle acceleration occurs in the jet of AGNs, it is reasonable to assume that a fraction of the jet kinetic power will be converted into UHECR kinetic energy~\citep{Eichmann_2018,Eichmann_2019,Matthews_Taylor_2021}. 
The intrinsic $\gamma$-ray luminosity is significantly correlated with the jet power of AGN from different categories~\citep{doi:10.1126/science.1227416,10.1093/mnras/stad065}. 
In addition, $\gamma$-ray emission is linked to particle acceleration and interactions in its neighborhoods~\citep[e.g.,][]{10.1093/mnras/stu088,Murase_Zhang_CenA_2023}.
In this way, using $\gamma$-ray luminosity as a possible normalization for the UHECR flux from different sources is well justified. Nevertheless, when considering the $\gamma$ rays emitted from particles accelerated in relativistic blobs moving along the jet, the $\gamma$-ray flux suffers a Doppler boost in the jet direction, while the same is unlikely for UHECR emission.

In this section, we explore the implications of using the observed and intrinsic $\gamma$-ray luminosity of sources as a proxy for the UHECR luminosity. 
As a case study, we use the $\gamma$AGN catalog selected from the analysis of the Pierre Auger Collaboration~\citep{Abdul_Halim_2024}. 
The selection contains jetted AGNs measured with the Fermi-LAT with a $\gamma$-ray flux $>3.3 \times 10^{-11}~\text{cm}^{-2}~\text{s}^{-1}$ between 10 GeV and 1 TeV on the 3FHL catalog of Fermi~\citep{Ajello_2017}. 
The determination of $\mathcal{D}$ is described in Appendix \ref{app:doppler}, and the relevant properties are in Table~\ref{tab:AGNs.prop}.

\subsection{$\gamma$-ray luminosity as proxy for UHECR luminosity}~\label{sec:gamma_proxies}

Assuming that the UHECR emission scales with the intrinsic characteristics of the source, but it is not beamed, the use of $L_\gamma^{obs}$ as a proxy for $L_{\rm{CR}}$ may overestimate the UHECR luminosity by a factor $\sim \mathcal{D}^q$, where $q$ take into account the different proxies between $\gamma$ rays and UHECR. The beaming effect becomes especially important when different classes of AGNs are included in the same analysis, such as radio galaxies and blazars. Due to the viewing angle, radio galaxies (RG) have a mean Doppler factor $\mathcal{D}_{\rm{RG}} \sim 2.6$~\citep{Ye_2023}, while for BL Lacs (BLL), $\mathcal{D}_{\rm{BLL}} \sim 10$~\citep{Zhang_2020,Ye_2023}.
Two scenarios are studied, both assuming an isotropic emission of UHECR. Firstly, the UHECR luminosity is assumed proportional to the intrinsic $\gamma$-ray luminosity of the source, and then the usual bolometric correction $q=4$ applies. Secondly, the UHECR luminosity is considered to be dependent directly on the radiative jet power, and then $q=2$\footnote{This value is valid for $\mathcal{D} \approx \Gamma$.}~\citep{Maraschi_2003,Chen_2023_Prad,doi:10.1126/science.1227416}. Since the observed luminosity is proportional to $\mathcal{D}^{q}$, this overestimates the UHECR flux from BLL when compared to RG on average by a factor of $\big( \nicefrac{\mathcal{D}_{\rm{BLL}}}{\mathcal{D}_{\rm{RG}}} \big)^q \sim 15-200$.

\begin{figure}
    \centering
    \includegraphics[width=0.5\linewidth]{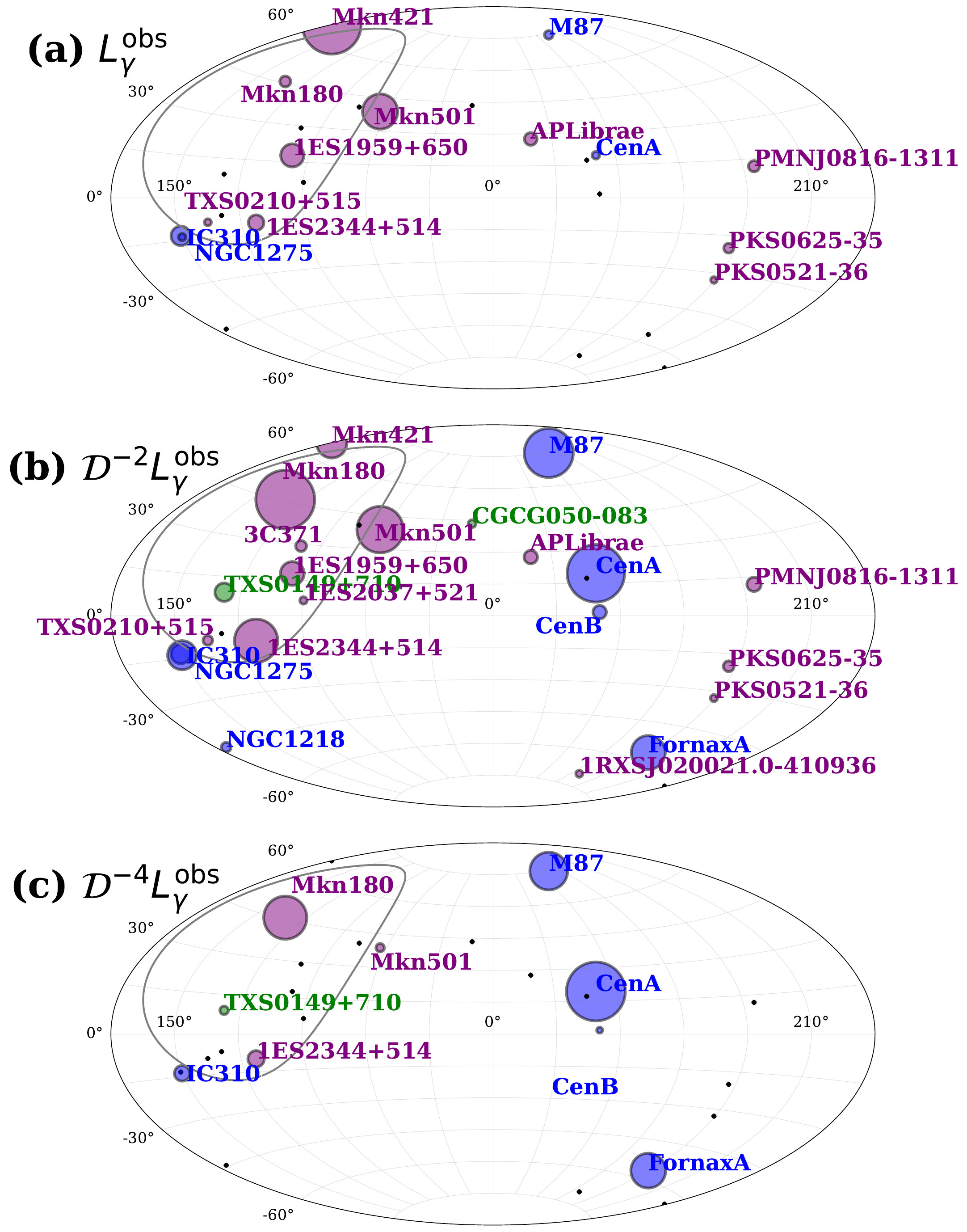}
    \caption{Luminosity weights used as a proxy for the UHECR flux of AGNs. Using (a) the observed $\gamma$-ray flux, (b) $\mathcal{D}^{-2} L_{\gamma}^{obs}$, and (c) $\mathcal{D}^{-4}L_{\gamma}^{obs}$. The circle size is linearly proportional to the contribution of each source, normalized by the contribution of the brighter source in each panel. Sources with a contribution above $10^{-2}$ of the maximum contribution are shown as circles (blue = RG, green = BCU, purple = BLL). Other sources are represented by black diamonds.}
    \label{fig:flux_map}
\end{figure}

Figure~\ref{fig:flux_map} shows the different luminosity weights used as proxies for the UHECR luminosity. 
The circle area represents the expected flux on Earth and is linearly proportional to the luminosity of each source. 
Sources with $L_{CR}$ smaller than $1\%$ of the largest value in each panel are shown as black diamonds of fixed size. 
When comparing panels (a) with (b) and (c), it is clear the importance of considering the beaming effect in the $\gamma$-ray luminosity of each source. 
Even though Mkn~421 is the dominant source using $L^{\rm{obs}}_\gamma$ (case a), it becomes negligible in any scenario where Doppler boosting effects are considered (cases b and c). 
In cases where the intrinsic source characteristics are considered (cases b and c), the radio galaxies Cen~A, M87, and Fornax~A are the brightest sources in the field of view of the Pierre Auger Observatory. Considering all the sky, the distant ($\sim200$~Mpc) blazars Mkn~180 can dominate over the nearby radio galaxies. This blazar presents a small Doppler factor ($\sim1.4$) when compared to the others BLL in the sample. Note that the strong dependence on $\mathcal{D}$ makes the estimation of the flux very sensitive to its uncertainties: comparing the $\gamma$-ray flux from CenA and Mkn180, a Doppler factor $\mathcal{D}_{Mkn180} \sim \big( \nicefrac{ F_{3fhl}^{Mkn180} }{ F_{3fhl}^{CenA} } \big)^{1/q} \sim 1.2-1.4$ is enough to make the UHECR luminosity of Mkn180 equal to that from CenA.

\vspace{1cm}
\subsection{Results from the combined fit}

The proxies for the cosmic ray luminosity only provide an estimation of the emissivities of each source considered. 
UHECR undergo different energy and primary-dependent energy losses during propagation, which modulate the final spectrum and composition. For that reason, the final contribution from each source will strongly rely on the astrophysical model assumed for the sources, i.e., their injected spectra and compositions. 
To take this into account, we have performed a combined spectrum and composition fit, following the approach of previous works by the Pierre Auger Collaboration~\citep{Aab_2017,Abdul_Halim_2024}. 
1-D simulations in \texttt{CRPropa3}~\citep{Batista_2016} were performed in a uniform grid of energy (from $10^{18}$~eV to $10^{22}$~eV with 10 bins per decade), and distance (from $3$ to 3342 Mpc in 118 bins in log) for each of the five representative primaries, $^{1}$H, $^{4}$He, $^{14}$N, $^{28}$Si, and $^{56}$Fe. 
All the energy losses and the EBL model of \cite{Gilmore_2012} were considered. 
A smearing was introduced in the arrival directions via a Von Mises-Fisher distribution~\citep{1953RSPSA.217..295F} similar to that of ~\citet{Abdul_Halim_2024} (eq. 2.14), with $\Delta_0 = 5^{\circ}$ and $R_0 = 10$~EV. 
The direction-dependent exposure of the experiment was taken into account to consider only events with zenith angles smaller than 60 degrees. 
The effective spectrum of each source was fitted to

\begin{equation}
dN/dE (E) = 
\begin{cases}
N_{s} F_{i} E^{-\Gamma}, \rm{ \ \ \ \ \ \ \ \ \ \ \ \ \ \ \ \ \ \ \ \, \ for \ } E \leq Z_i R_{\rm{max}} \\
N_{s} F_{i} E^{-\Gamma} e^{(1-E/(Z_i R_{\rm{max}}))}, \rm{ \ for \ } E > Z_i R_{\rm{max}}
\end{cases},
\end{equation}
where the free parameters of the fit are the spectral index, $\Gamma$, the maximum rigidity at the sources, $R_{\rm{max}}$, the main normalization, and the contribution of each species, $F_i$. 
$F_i$ is defined as the total contribution of a primary between 1 EeV and the corresponding maximum energy, $ZR_{\rm{max}}$. 
This definition relates to that of \cite{Aab_2017}, $f_i$ (i.e., the relative contribution of each primary for an energy bin below the maximum energy of protons) as $f_i = F_i / (Z_i R_{\rm{max}})^{(\Gamma-1)}$.
Such quantity was chosen because it provides a more robust minimization procedure for the fit. 
The sources were divided into two classes: a homogeneous distribution of background sources with equal emissivity and no source evolution and the so-called local sources, whose individual emissivities are modulated by the values in Table~\ref{tab:AGNs.prop}.
The relative contribution of local and background sources is given by the parameter $\alpha = J_{\rm{local}} (E=10^{19.5} \, \rm{eV}) / J_{\rm{background}} (E=10^{19.5} \, \rm{eV})$. Both the normalization for the background sources and $\alpha$ are also free parameters of the fit.

The simulations are multiplied by the weights described above and compared to the spectral data above $10^{18.7}$~eV from \cite{Verzi:2020opp} and measurements of the first and second moments of the $X_{\rm{max}}$ distributions from \cite{Yushkov:2020nhr}\footnote{Taken from Pierre Auger's public dataset: \url{https://www.auger.org/science/public-data/data}}. The fit was performed by minimizing the $\chi^2$ distance between the model and the spectral/composition data.

The systematic uncertainties in the spectrum and $X_{\rm{max}}$ were addressed by performing a new fit with the data shifted by $\pm 14\%$ for the energy and $\pm 1\sigma$ for the $X_{\rm{max}} (E)$ according to \cite{PierreAuger:2014sui}. 
Resulting in a total of 9 fits for each of the cases (a), (b), and (c). 
No shift in $X_{\rm{max}}$ was preferred by the fit for every case. For energy, on the other hand, the data was best described with $E \rightarrow E - 14\%$, $E \rightarrow E + 0\%$, and $E \rightarrow E + 14\%$ for $L^{\rm{obs}}_\gamma$, $\mathcal{D}^{-2} L^{\rm{obs}}_\gamma$, and $\mathcal{D}^{-4} L^{\rm{obs}}_\gamma$ respectively.
The $X_{\rm{max}}$ moments were related to the mass compositions using the EPOS-LHC hadronic model~\citep{epos}. Different hadronic models were tested, but the hierarchy between the fit results for each proxy remained the same.

\begin{figure}
    \centering
    \includegraphics[width=0.45\linewidth]{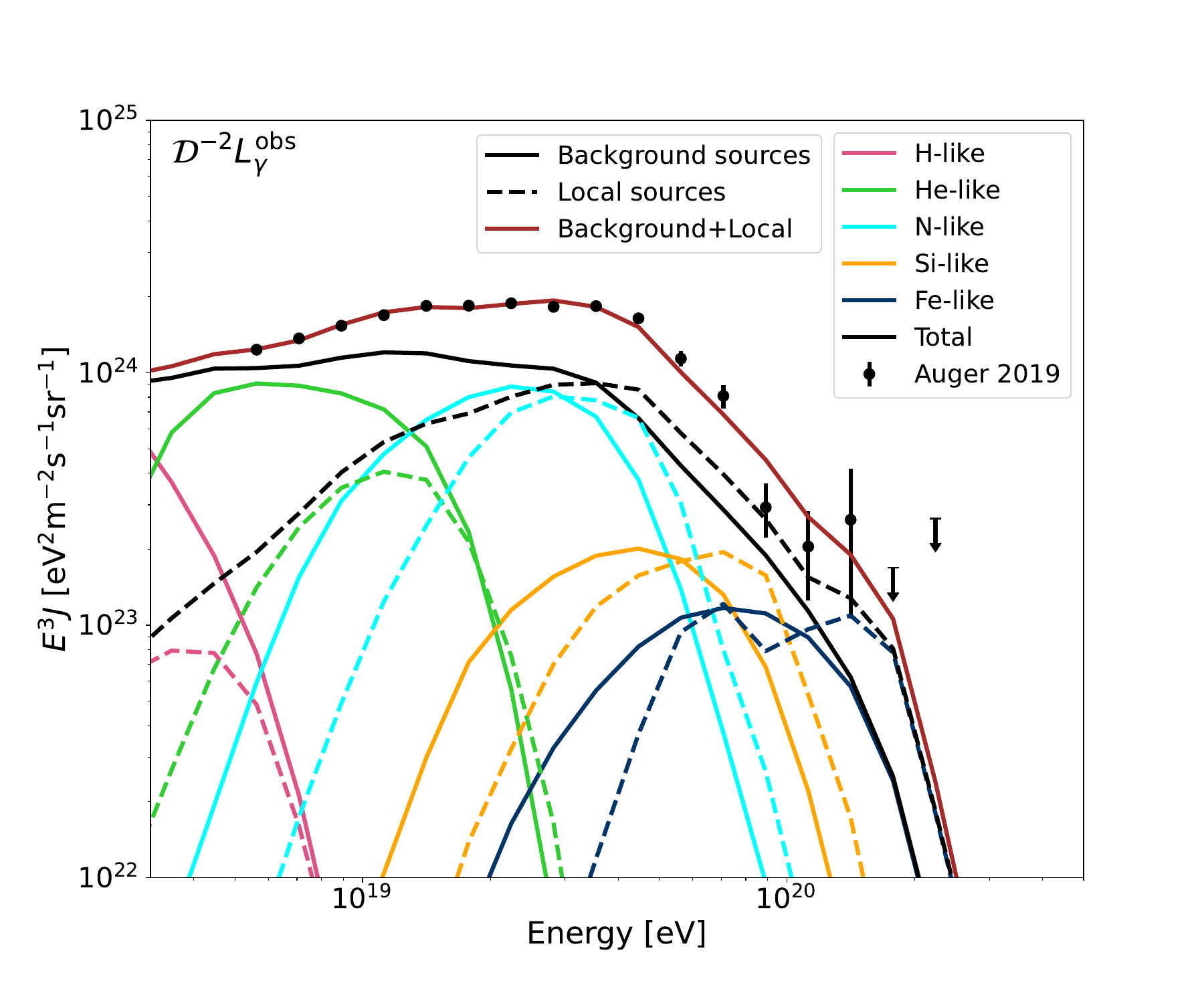}
    \includegraphics[width=0.45\linewidth]{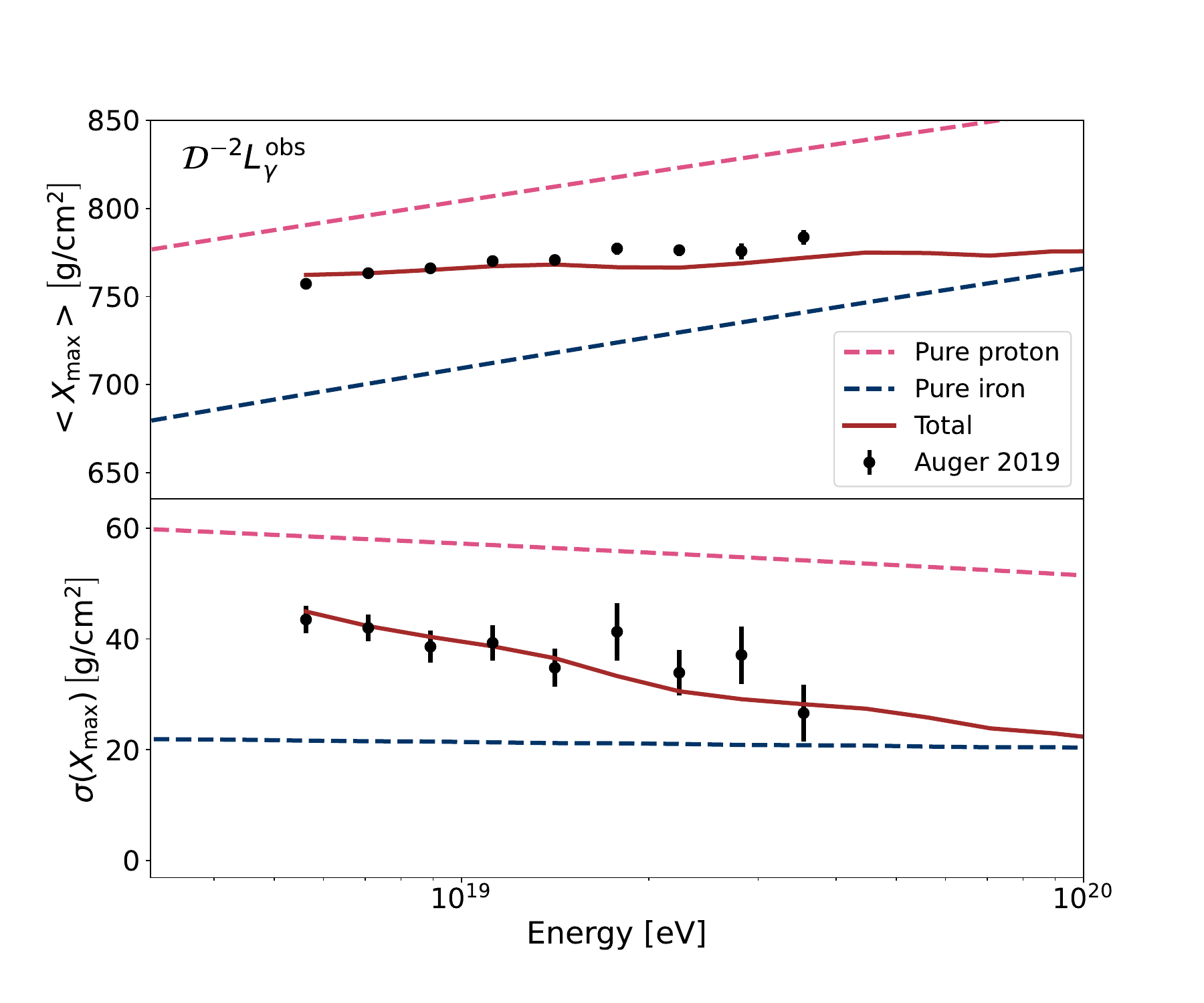}
    \caption{Spectrum (left) and first two moments of the $X_{\rm{max}}$ distribution (right) for the best-fit case scenario using $D^{-2} L_{\gamma}^{\rm{obs}}$ as a proxy. The full and dashed lines in the left panel show the contribution of background and local sources, respectively. The different masses arriving on Earth are grouped into H-like ($A=1$), He-like ($2\leq A \leq 4$), N-like ($5 \leq A \leq 22$), Si-like ($23 \leq A \leq 38$), and Fe-like ($39 \leq A \leq 56$) and are represented by the different colors. The dashed lines on the right panel show the expectations for an extreme pure composition scenario. The best-fit parameters are $\Gamma = -2.0^{+0.06}_{-0.61}$, $\log_{10} \left( R_{\rm{max}} / \rm{V} \right) = 18.11^{0.02}_{-0.04}$, and $\alpha = 49 \pm 1\%$. No systematic shift for energy or $X_{\rm{max}}$ is preferred.}
    
    \label{fig:spectrum}
\end{figure}

Figure~\ref{fig:spectrum} shows the spectrum and $X_{\rm{max}}$ moments for the best fit of the $\mathcal{D}^{-2} L_{\gamma}^{\rm{obs}}$ scenario. Similarly to the previous results found by the Pierre Auger Collaboration in the absence of an extragalactic magnetic field, a very hard spectrum with $\Gamma < 0$ with a strict rigidity cutoff, $R_{\rm{max}} = 10^{18.11}$~V, is preferred.

The resulting goodness of fit were $\left(\chi^2/\rm{NDF}\right)_{L^{\rm{obs}}_\gamma} = 114.9/25 \approx 4.6$, $\left(\chi^2/\rm{NDF}\right)_{\mathcal{D}^{-2} L^{\rm{obs}}_\gamma} = 77.6/25 \approx 3.1$, and $\left(\chi^2/\rm{NDF}\right)_{\mathcal{D}^{-4} L^{\rm{obs}}_\gamma} = 98.6/25 \approx 3.9$, demonstrating an improvement of the fit when the new proposed proxies are considered, in special the $\mathcal{D}^{-2} L^{\rm{obs}}_\gamma$ case. The main reason for that comes from the relative contribution of farther local sources. A larger contribution from a very local source such as Cen~A will result in a total spectrum with a spectral index very similar to the intrinsic spectral index assumed for the sources.
When larger relative contributions from farther sources are predicted, an effective spectral index is seen in the combined local source flux. 
This is mostly due to two effects: the suppression will occur at different energies for sources at different distances, and more UHECR from farther sources will go through photodisintegration, resulting in a larger contribution to lower energies in the spectrum. This shapes the final spectrum on Earth and vastly increases the contribution of local sources to the lower energy end of the spectrum, leading to a poor fit of the data.

\begin{figure}
    \centering
    \includegraphics[width=0.48\linewidth]{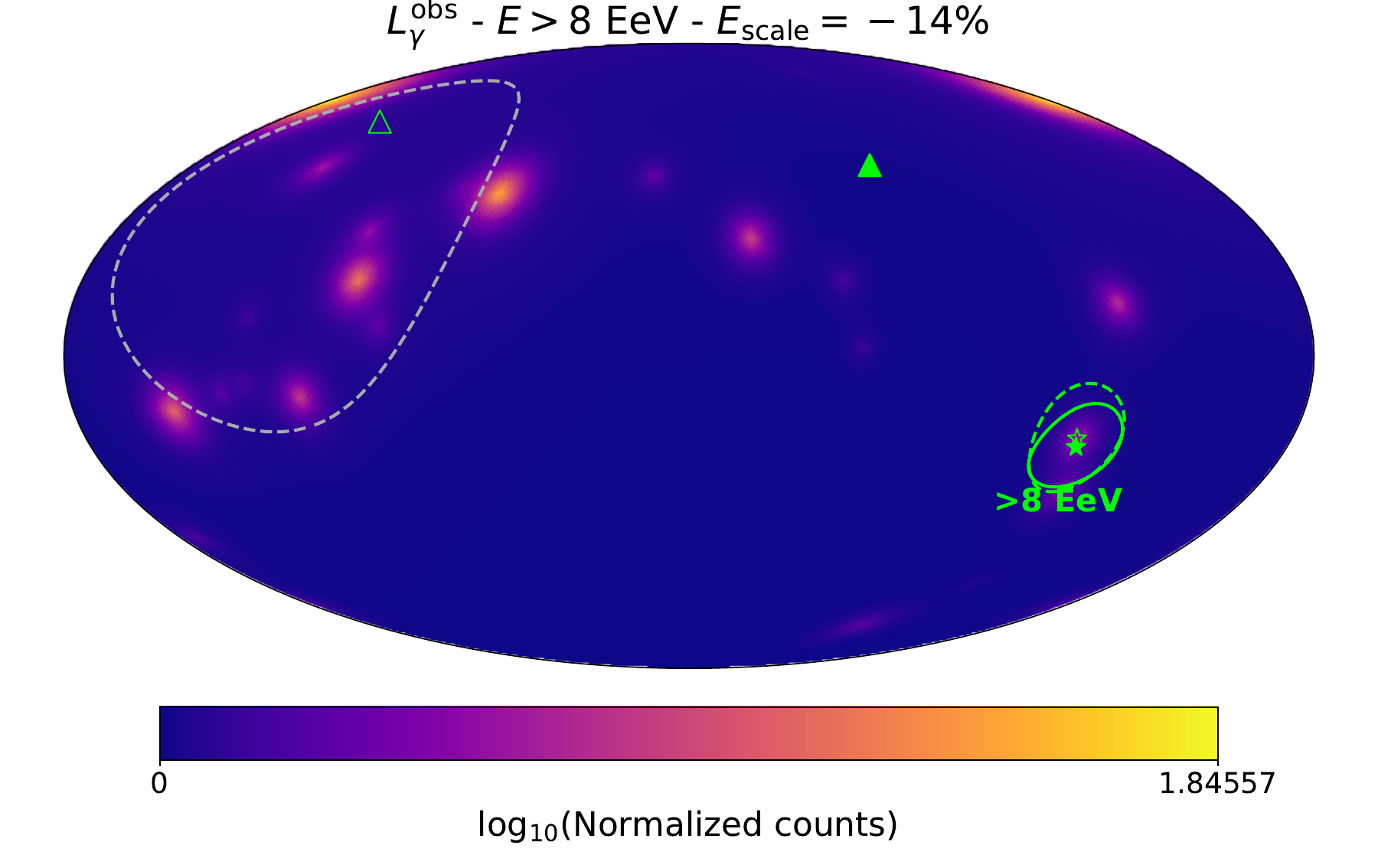}
    \includegraphics[width=0.48\linewidth]{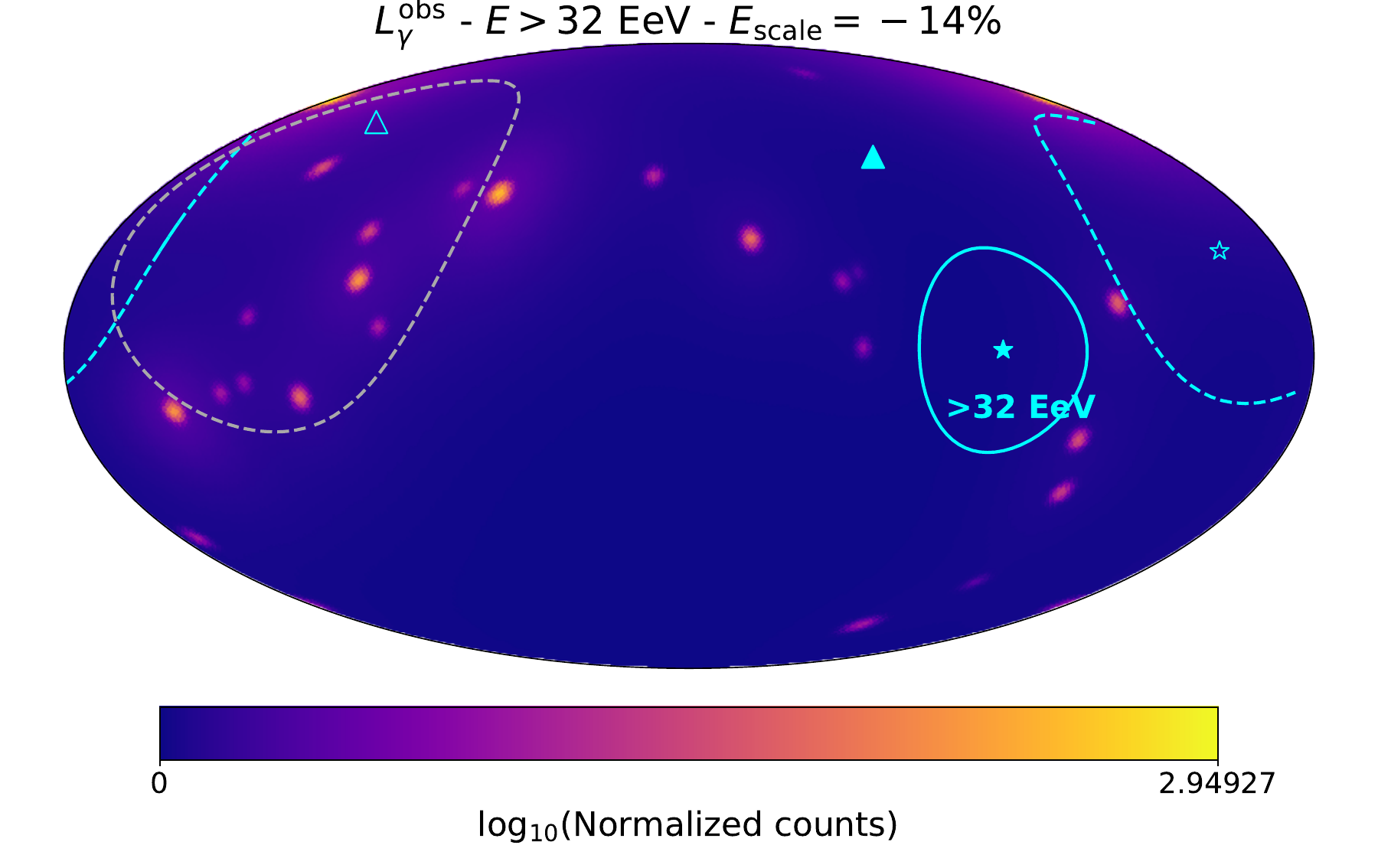}
    \includegraphics[width=0.48\linewidth]{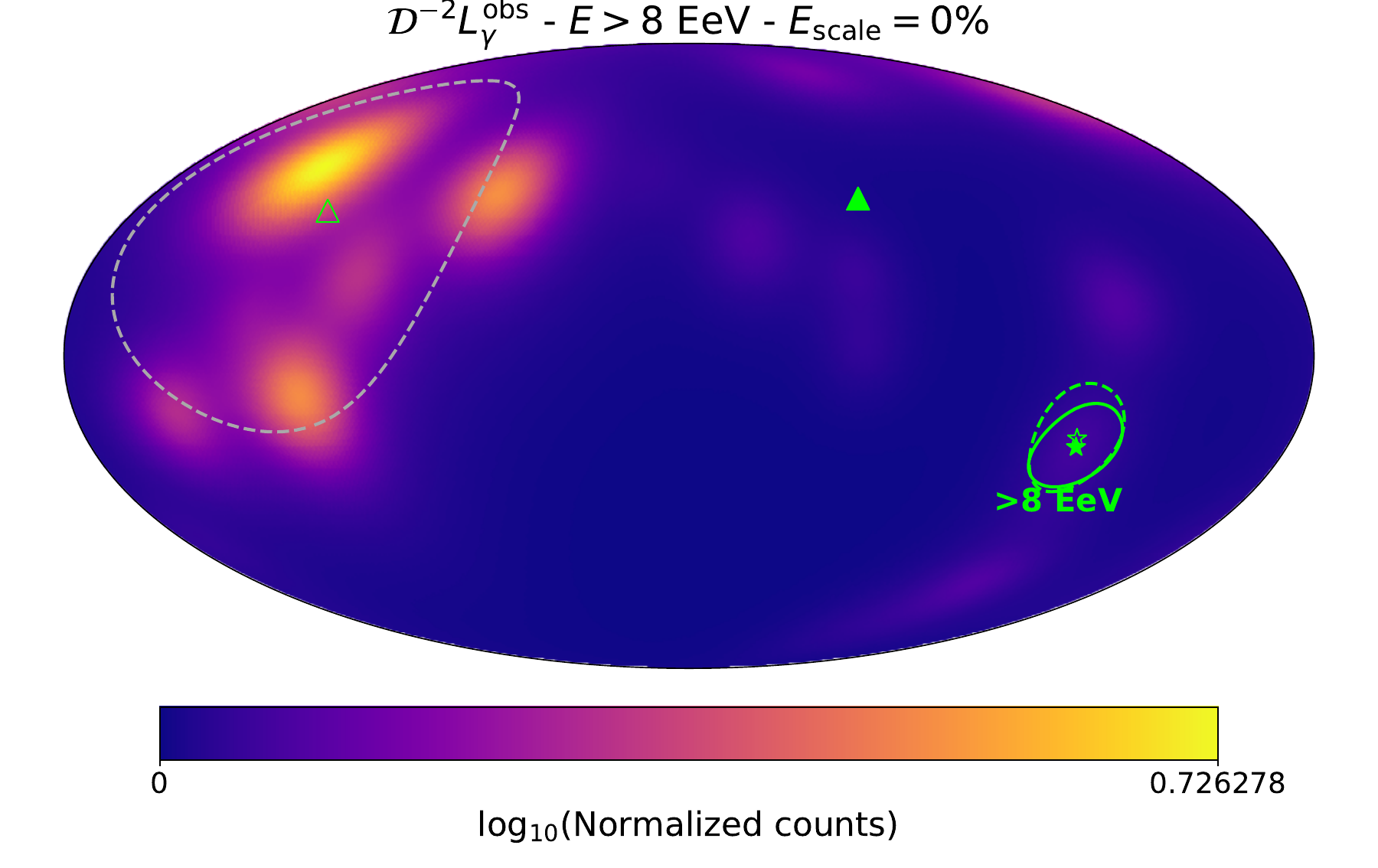}
    \includegraphics[width=0.48\linewidth]{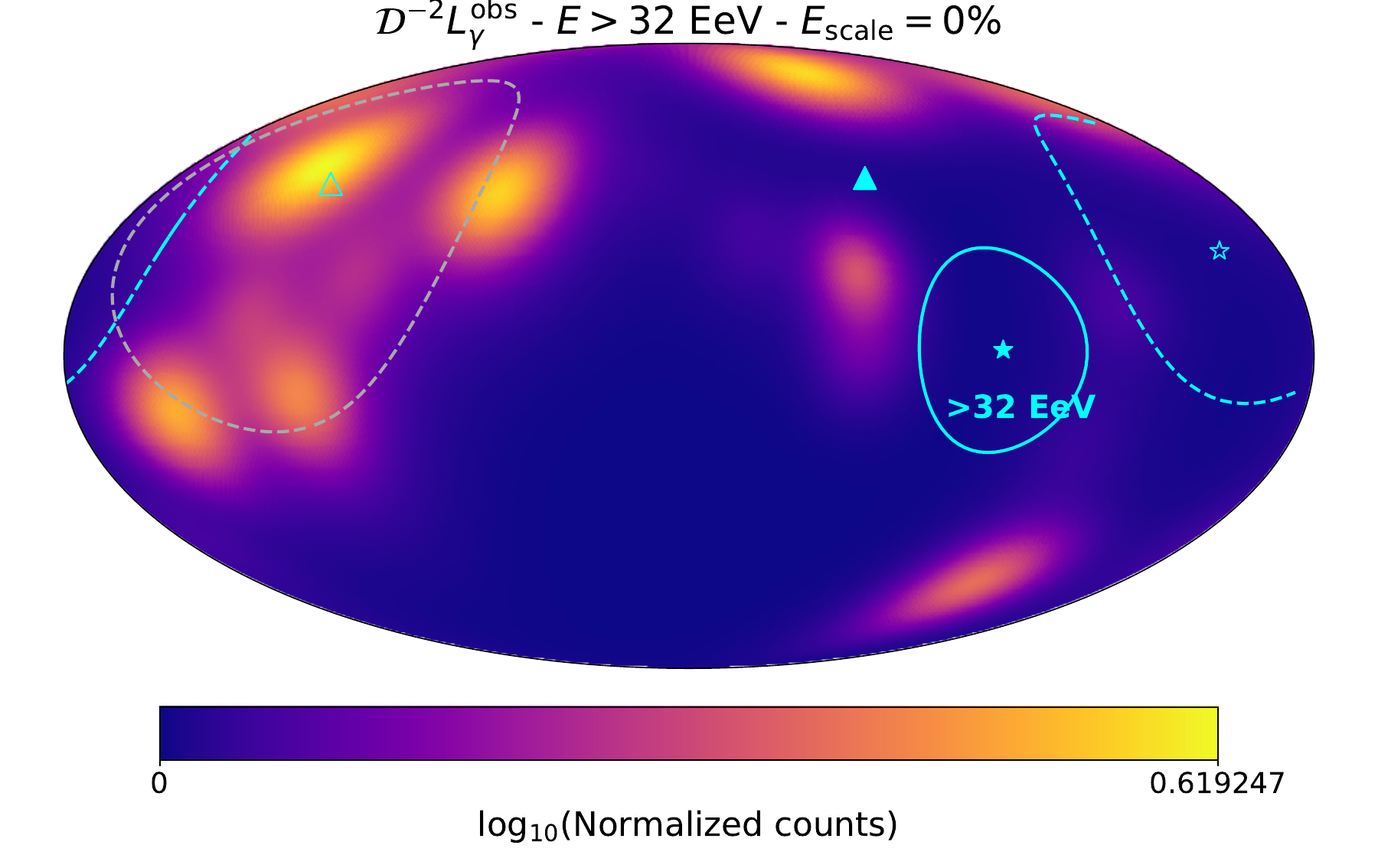}
    \includegraphics[width=0.48\linewidth]{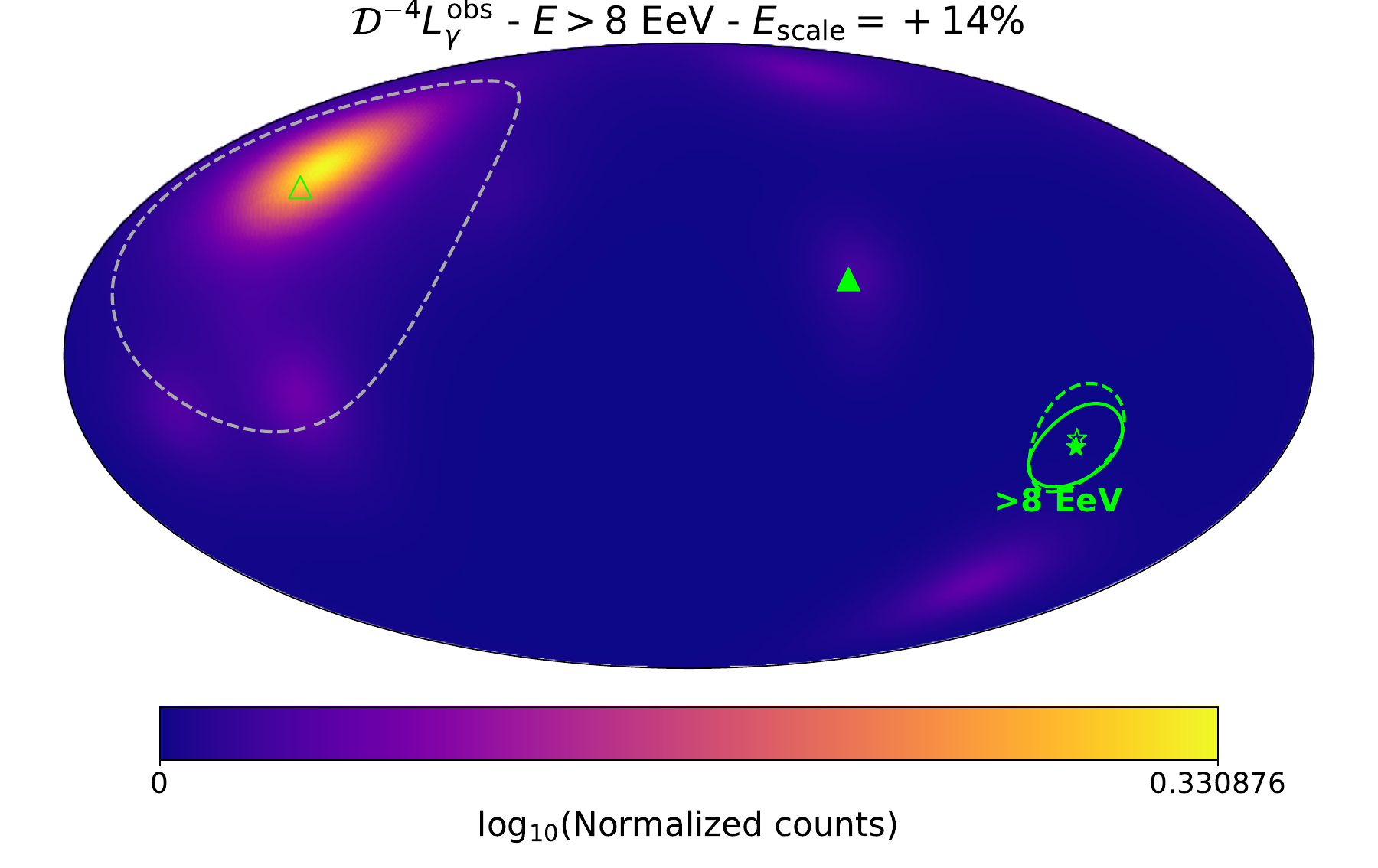}
    \includegraphics[width=0.48\linewidth]{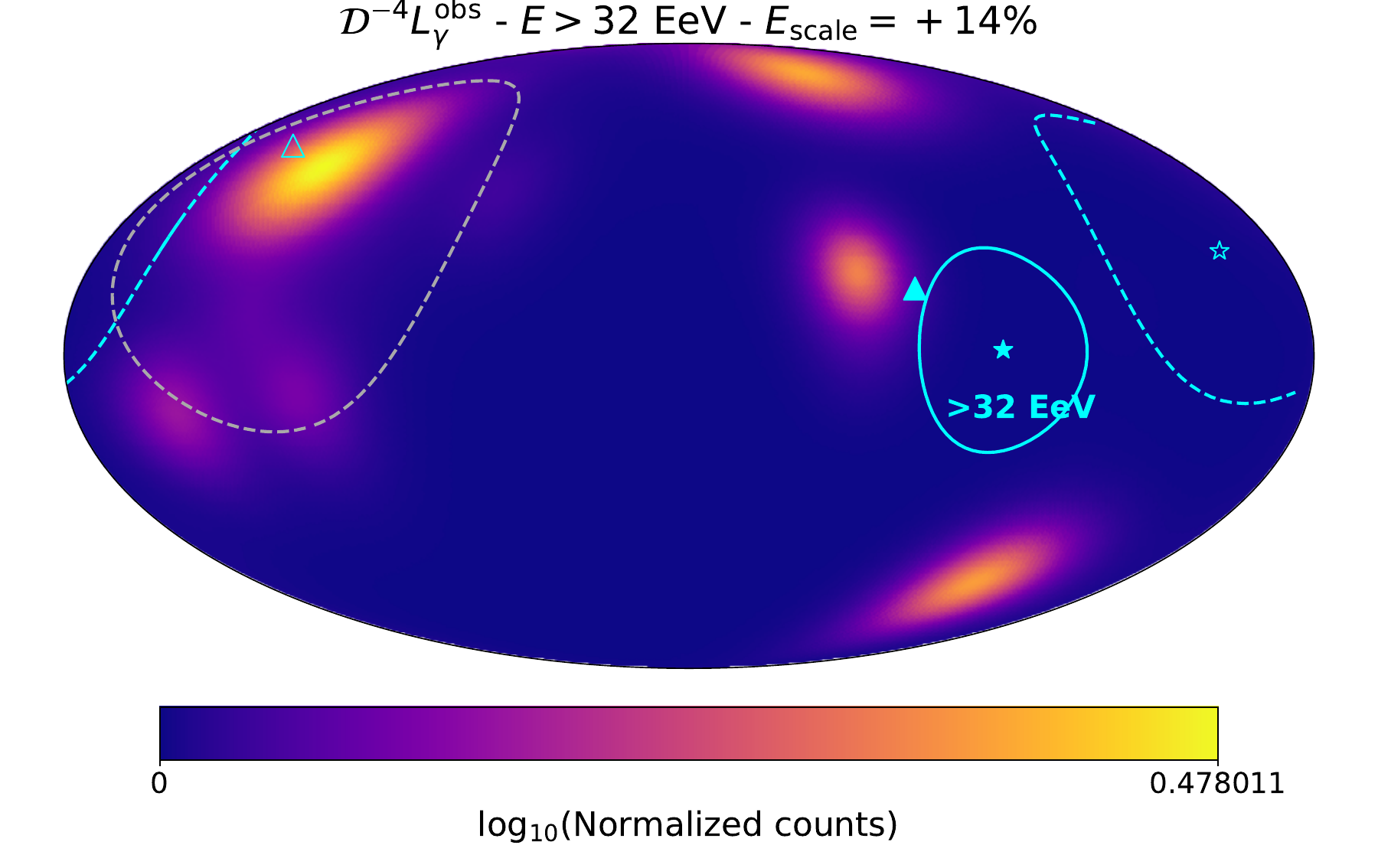}
    \caption{Arrival direction maps. Each row shows the results from the best-fit scenario considering each of the $\gamma$-ray proxies and the left and right columns show the result for $E> 8$~EeV and $E> 32$~EeV respectively. The counts are normalized to the bin with the fewest counts, i.e., regions that have negligible contribution from local sources. The lime and cyan full stars and contours show the dipole measured by the Pierre Auger Observatory for $E>8$~eV and $E>32$~eV, while the lime and cyan empty stars and dashed contours show the full-sky dipole measured by a combined analysis of the Pierre Auger Observatory and Telescope Array~\citep{PierreAuger:2023mvf}. Open and closed triangles show the obtained dipoles for the full sky and just for the Auger field of view for our scenarios, respectively. The dashed gray contour shows Auger's field of view considering events with zenith angles smaller than $80^{\circ}$. A rigidity-dependent blurring following a Von Mises-Fisher~\citep[eq. 2.14]{Abdul_Halim_2024} with $\Delta_0 = 5^{\circ}$ and $R_0 = 10$~EV was used for the local sources.}
    \label{fig:maps}
\end{figure}

\begin{figure}
    \centering
    \includegraphics[width=0.85\linewidth]{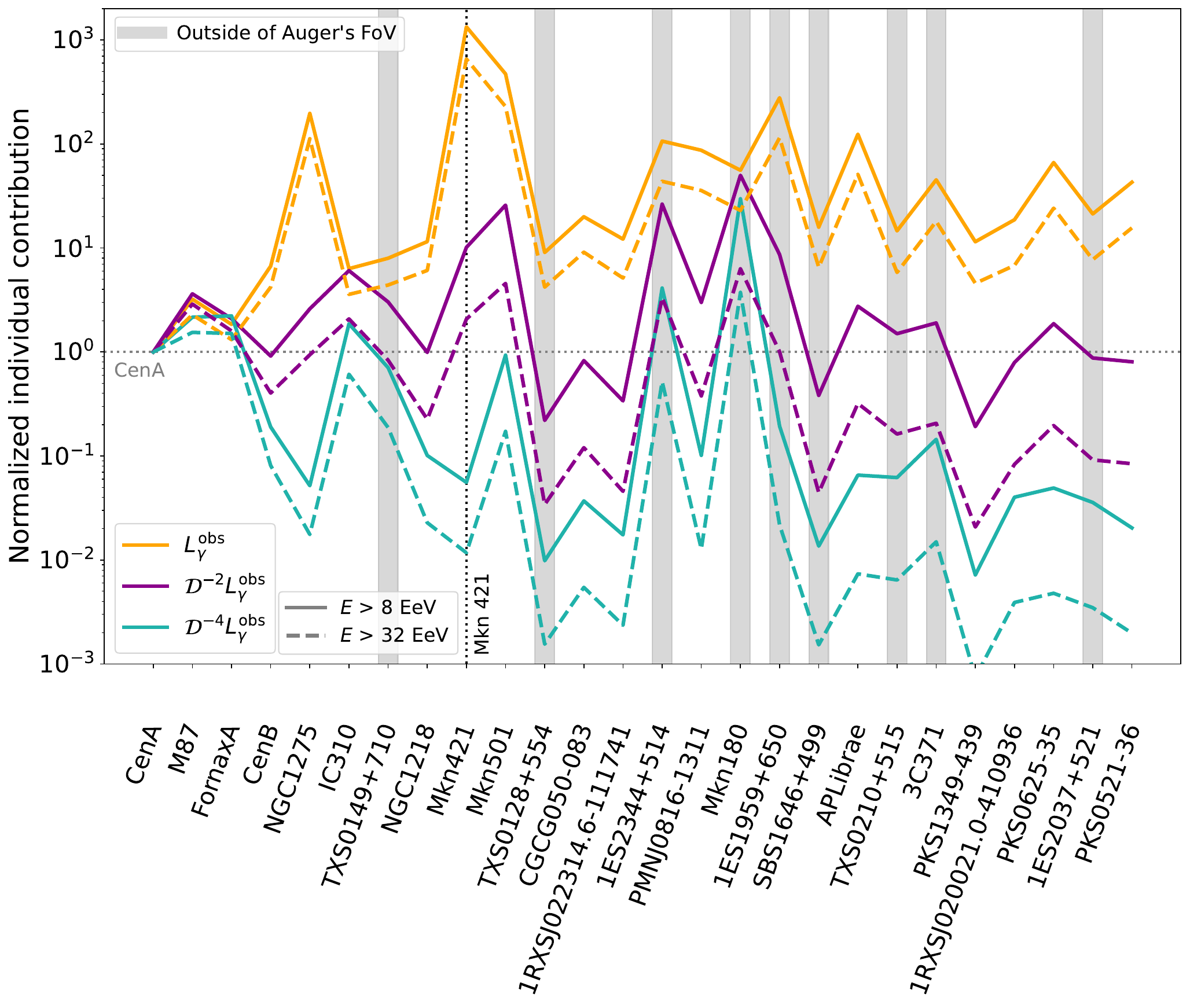}
    \caption{Relative contribution to the spectrum from each individual source. The contribution from Cen A is taken as the reference for each case. Different line colors show the different proxies considered in this work, while continuous and dashed lines show the contributions for energies above 8 and 32 EeV, respectively. Sources outside of the field of view of the Pierre Auger Observatory are highlighted by a gray band.}
    \label{fig:spectrum2}
\end{figure}

The most important effect is seen in the distributions of arrival directions. 
Figure~\ref{fig:maps} shows the arrival direction maps for $E > 8$~EeV and $E > 32$~EeV. The counts for each of the 49152
 pixels with equal solid angle are normalized to the pixel with the fewest counts, such that pixels with only contribution from background sources are set to 1. An overall normalization is arbitrary as it does not influence the structures of the map. A log scale is chosen to help highlighting different hotspots. The reconstructed dipole directions for the full sky and field of view of the Pierre Auger Observatory are also shown for each case. Figure~\ref{fig:spectrum2} shows the relative contribution of each local source to the spectrum for the same energy ranges. 

The hotspots change for each scenario considered. The dominance of Mkn~421 seen in $L^{\rm{obs}}_{\gamma}$ vanishes for the new proxies. As seen in figure~\ref{fig:spectrum2}, its contribution is suppressed by 2 to 5 orders of magnitude depending on the proxy and energy range. This source was reported in~\cite{Abdul_Halim_2024} as the biggest challenge for performing a combined fit of the Pierre Auger Collaboration data using arrival direction, due to its relative large distance, position within Auger's field of view, and predicted UHECR contribution which was in contrast with data. This challenge seems to be overcome by the new proxies proposed in the work.

For both the $\mathcal{D}^{-2} L^{\rm{obs}}_{\gamma}$ and $\mathcal{D}^{-4} L^{\rm{obs}}_{\gamma}$ cases, the main hotspots are seen around Cen~A, Fornax~A, M87 and a few sources outside Auger's field of view, in particular Mkn~180 and 1ES2344+514. The lack of a hotspot near M87 in data could be explained by a shadowing due to the effect of extragalactic and galactic magnetic fiels~\citep{deOliveira_2022,Condorelli_2023, deOliveira_2023}. The remaining three hotspots are in relative agreement with the two hotspots reported by the Pierre Auger Observatory~\citep[$(\ell,b)\approx(310^\circ,20^\circ)$ and $(\ell,b)\approx(270^\circ,-75^\circ)$, ][]{Aab_2018_indication} and the hotspot reported by the Telescope Array Observatory~\citep[$(\alpha,\delta)=(144^\circ,40.5^\circ)$, or, in galactic coordinates, $(\ell,b)=(181.5^\circ,47.8^\circ)$, ][]{Kim:2023ksw}.

The reconstructed dipoles are also improved with the new proxy assumptions. The reconstructed amplitude for the $L^{\rm{obs}}_{\gamma}$ proxy and $E > 8$~EeV was $d = 14 \pm 1$\%, too large when compared to the amplitude of $7.4^{+1.0}_{-0.8}$\% reported by~\cite{Abdul_Halim_2024_19yr}. This again comes from an overcontribution of Mkn~421. For the new proxies, $\mathcal{D}^{-2} L^{\rm{obs}}_{\gamma}$ and $\mathcal{D}^{-4} L^{\rm{obs}}_{\gamma}$, the suppression of Mkn~421 leads to smaller reconstructed amplitudes of $d = 7.5 \pm 0.2$\% and $d = 4.1 \pm 0.5$\%. Higher order multipoles are not investigated in this work, as they are significantly more susceptible to diffusion in the galactic and extra-galactic magnetic fields, neglected in this work~\citep{deOliveira:2023kvu}. 

Improvements can also be seen in the reconstructed direction of the dipole. As seen in figure~\ref{fig:maps}, none of the reconstructed dipole directions agree within 1$\sigma$ with the experimental data. Nevertheless, for the Auger dipole, the tension between the directions of the observed and modeled dipoles is reduced with the new proxies. We calculate the angular distance between the reconstructed and measured dipole and compare that to the 1$\sigma$ uncertainty in the measurement and shown that for the field of view of Auger an improvement from 5.9 (2.1)$\sigma$ to 3.5 (1.1)$\sigma$ is found for $E> 8 (32)$~EeV. The results for both Auger's field of view and full sky are shown in Table~\ref{tab:dipole}. Nevertheless, it is worth pointing out that the spectral and composition fit were done considering only Auger data. Different results would be expected if data from Auger and Telescope Array would have been combined as is the case of the full-sky measurements.

\begin{table}[]
    \centering
    \caption{Significance of the tension of the predicted direction of the dipole with experimental data.} 
    \label{tab:dipole}

\begin{tabular}{c|c|c|c|c}
    \hline\hline
Field of view & Energy & $L^{\rm{obs}}_{\gamma}$ & $\mathcal{D}^{-2} L^{\rm{obs}}_{\gamma}$ & $\mathcal{D}^{-4} L^{\rm{obs}}_{\gamma}$ \\
    \hline
  \multirow[c]{2}{*}{Auger} & $> 8$ EeV & 5.9$\sigma$ & 4.9$\sigma$ & 3.5$\sigma$ \\
   & $> 32$ EeV & 2.1$\sigma$ & 2.0$\sigma$ & 1.1$\sigma$ \\
   \hline
  \multirow[c]{2}{*}{Full sky} & $> 8$ EeV & 8.7$\sigma$ & 9.8$\sigma$ & 8.7$\sigma$ \\
   & $> 32$ EeV & 1.5$\sigma$ & 1.7$\sigma$ & 1.2$\sigma$ \\
     \hline\hline
\end{tabular}
\end{table}

\section{Summary and Discussion}
\label{sec:conclusions}

AGN are among the main candidates to accelerate cosmic rays up to $10^{20}$~eV. Different views on the relative contribution from each source have been explored in previous works, leading to a significant change in the agreement with data. 
In this work, we conciliate these studies with a thorough investigation of the $\gamma$-ray luminosity proxy for UHECR.

\begin{itemize}
    \item \textbf{The association between $\gamma$ rays and UHECR is weak, but not impossible}: $\gamma$ rays can have an origin in leptonic or hadronic scenarios. In the hadronic case, the cosmic ray energy necessary to produce a $\sim$TeV $\gamma$ ray is $\sim$PeV, considerably below the UHECR regime. 
    $\gamma$ rays emitted by EeV UHECR will likely have energies not accessible by $\gamma$-ray observatories due to EBL and CMB attenuation. In this way, the correlation between $\gamma$ rays and UHECR can be considered weak, although it should not be ignored, since the detection of $\gamma$ radiation implies the existence of regions where the acceleration of charged particles occurs. Furthermore, the $\gamma$-ray luminosity is related to the jet power in AGN, which can be related to its UHECR luminosity;
    \item \textbf{The use of the observed flux of $\gamma$ rays as proxies for UHECR implicitly assume that both are beamed}: When the $\gamma$-ray flux is used as a proxy for the UHECR flux, there is an \textbf{implicit hypothesis} that UHECRs are subject to the same beaming effect of $\gamma$ rays. It is unclear if UHECRs are accelerated in relativistic blobs as is assumed for $\gamma$ rays. However, even in that case, the magnetic fields in the acceleration regions, jets, and lobes will likely decollimate the UHECR beam, as shown in numerical studies. This way, the expected emission cone of UHECR is larger than that of $\gamma$ rays;
    \item \textbf{The correction of the observed flux of $\gamma$ rays as proxies for UHECR is source dependent and on average decreases the contribution for farther sources:} Assuming that UHECRs are not beamed as $\gamma$ rays, the intrinsic $\gamma$-ray luminosity or the radiative jet power are better proxies than the observed $\gamma$-ray luminosity. The relation between them with the observed $\gamma$-ray luminosity is given by $L_{\gamma}^{\rm{int}} = \mathcal{D}^{-q} L_{\gamma}^{\rm{obs}}$, with $q\sim2-4$. It becomes highly relevant when different AGN classes are considered, such as blazars ($\mathcal{D} \sim 10$) and radio galaxies ($\mathcal{D} \sim 2$);
    \item \textbf{Using intrinsic $\gamma$-ray luminosity or the radiative jet power as UHECR proxy gives a better fit to the Pierre Auger Observatory data:} A combined fit of spectrum and composition data performs better when $L_{\gamma}^{\rm{int}}$ is considered. The spectral shape of local sources is changed due to an increase in the relative contribution of closer sources. For the simple combined fit proposed in this work, an improvement from $\chi^2/\rm{NDF} = 4.6$ to $\chi^2/\rm{NDF} = 3.1$ is found.
    \item \textbf{Using intrinsic $\gamma$-ray luminosity or the radiative jet power as UHECR proxy conciliates the arrival directions data:} The relative contribution of each source, particularly at the highest energies, is changed for different proxy assumptions. When $L_{\gamma}^{\rm{int}}$ is considered, the strong expected contribution from Mkn~421 vanishes. The predicted dipole shifts from $5.9 \ (2.1)\sigma$ up to $3.5 \ (1.1)\sigma$ away from the one measured by the Pierre Auger Observatory for $E>8$~EeV ($>32$~EeV). The predicted hotspots also change significantly. For the new proxies proposed here, three main hotspots appear in locations similar to that of the two hotspots measured by the Pierre Auger Observatory and the hotspot measured by the Telescope Array Experiment.
\end{itemize}

The intrinsic $\gamma$-ray luminosity or the radiative jet power used here appears to be a better proxy than the observed $\gamma$-ray luminosity.
Since the intrinsic $\gamma$-ray luminosity is related to the jet kinetic power~\citep{10.1093/mnras/stad065}, it agrees with authors who argue that the UHECR luminosity must scale with the jet power~\citep{Eichmann_2018, Matthews_Taylor_2021}. In particular, \cite{Matthews_Taylor_2021} suggests that radio luminosity is a better proxy for $L_{\rm{CR}}$, with reservations due to UHECR transport and particular characteristics of different sources.

Our results resonate with other works that found AGN catalogs as promising source candidates~\citep{Eichmann_2018,Eichmann_2019,Eichmann_2022,deOliveira_2022,deOliveira_2023}. 
The contribution of the radio galaxies Cen~A and Fornax~A have been proposed as responsible for the dipole and hotspots measured by the Pierre Auger Observatory~\citep{matthews_fornax,deOliveira_2022}, and a powerful emission from Cen~A and Fornax~A from past enhanced activity cannot be ruled out~\citep{matthews_fornax}.

It is important to emphasize that since radio galaxies does not present the $\gamma$-ray signal enhanced by the relativistic beaming fewer of them are detectable with the increasing distance when compared to blazars. This selection bias can underestimate the overall contribution of radio galaxies concerning blazars, especially when distant sources are considered.

In contrast to a fully isotropic emission, \cite{rachen2019parameterizedcatalogradiogalaxies} propose that blazars could have an additional beamed UHECR emission. 
In this work, we neglect that based on the decollimation effect on the source region. 
Due to the energy dependence on the magnetic scattering of charged particles, the decollimation of a possible UHECR beam will also be energy-dependent. 
We do not account for this effect here, and it will be addressed in future works.

After leaving the source, the extragalactic and galactic magnetic fields also should decollimate an eventual residual UHECR beam. 
In this work, we follow the simplistic assumption of blurring the arrival directions due to a turbulent component of the extragalactic magnetic field. 
The regular component of the EGMF can cause amplification/suppression of sources and a shift of arrival directions, which can be significant even to nearby sources~\citep{lang2021,deOliveira_2022,deOliveira_2023} and must be taken into account in a detailed study. 
However, the EGMF structure seems to have a minor effect in the dipole direction for $E>32$~EeV if the flux is dominated by nearby radio galaxies~\citep{deOliveira_2023}. 
Deflections by the galactic magnetic field were not taken into account either. 
A detailed exploration of the effect of the cosmic magnetic fields is beyond the scope of this study, since the extragalactic and galactic magnetic fields are complex and the effects on UHECRs are still open questions~\citep{Bakalova_2023, Diego_Harari_2002, hackstein_2018, deOliveira_2022}.

Finally, the dominant sources of UHECR can only be addressed by quantitative comparison of scenarios aiming at reproducing energy spectrum, composition, and arrival directions. Other scenarios such as SBGs (especially due to the high rate of transient event, such as long gamma-ray bursts)~\citep{marafico2024closingnettransientsources}, and large-scale shocks in galaxy clusters and filaments~\citep{Simeon:20233q}, for example, remains as viable sources of UHECR~\citep{kachelriess2022extragalacticcosmicrays}.
The combined fit used in this work is a simplified version that does not take into account the full $X_{\rm{max}}$ distributions and the arrival direction distribution. 
Still, the results shown here for arrival direction maps and dipole indicate that a full fit with the proxies here proposed will decrease the tension with the data.

Therefore, even with the intrinsic limitations addressed above, the results of this work strengthen the hypothesis of AGN as candidates for the origin of UHECR and provide the community with a more robust hypothesis about the proxies for $L_{\rm{CR}}$ using $L_{\gamma}$. In addition, it potentially conciliates the results using different proxies, since both the radio and the intrinsic $\gamma$-ray luminosity scales with the jet power.
These assumptions could be used in future studies that aim to model the data or look for correlations in the arrival directions data from the Pierre Auger Observatory and the Telescope Array Experiment.

\section*{Acknowledgments}
The idea for this work was developed during a series of two joint FAPESP/BAYLAT workshops titled ``Astroparticle physics in the era of CTA and SWGO'' at the Friedrich-Alexander-Universität Erlangen-Nürnberg and the Instituto de Física de São Carlos, Universidade de São Paulo in 2023 and 2024. We acknowledge the generous support for these workshops by FAPESP (through grant number 2022/01271-7) and BAYLAT. The authors thank Vitor de Souza, James Matthews, and Teresa Bister for reading the paper and making useful comments. This study was financed, in part, by the São Paulo Research Foundation (FAPESP), Brasil. Process Number 2019/10151-2, 2020/15453-4, and 2021/01089-1. CO acknowledges the National Laboratory for Scientific Computing (LNCC/MCTI, Brazil) for providing HPC resources for the SDumont supercomputer (http://sdumont.lncc.br).

\bibliography{main.bib}

\appendix

\section{Doppler factors estimations} \label{app:doppler}

The energy flux integrated from $10$~GeV to $1$~TeV (erg~cm$^{-2}$~s$^{-1}$),
\begin{equation}
    S_\gamma^{0.01-1~\text{TeV}} = \int_{10~GeV}^{1~TeV} E \frac{dN}{dE} dE,
\end{equation}
measured by the Fermi-LAT satellite, was used to weigh the UHECR flux from each source. The \textbf{isotropic} $\gamma$-ray luminosity is estimated as
\begin{equation}
    L_{0.01-1~\text{TeV}}^{iso} = 4 \pi d^2 S_\gamma^{0.01-1~\text{TeV}}
\end{equation}
where $S_\gamma$ is the gamma-ray energy flux and $d$ is the distance, found in the Auger Catalogue~\citep{Abdul_Halim_2024}, and the redshift dependence is omitted since the most distant source is PKS0521-36, at $d \approx 241$~Mpc, whose $z \approx 0.056 \ll 1$.



The intrinsic and observed luminosities are related by the Doppler factor ($\mathcal{D}$) of the plasma in the jet by
\begin{equation}
    L^{int} = \mathcal{D}^q L^{iso},
\end{equation}
where $q=-2$ or $-4$ (see Section~\ref{sec:gamma_proxies}).

Obtaining reliable estimations of the viewing angle and bulk velocities is challenging, with different indirect methods proposed in the literature. These methods do not always agree~\citep{Liodakis_2017}. \citet{Liodakis_2018} obtained $\mathcal{D}$ based on the observed and equipartition brightness temperature of flares. \citet{Zhang_2020} determined $\mathcal{D}$ from correlations between the $\gamma$-ray luminosity and broad-line luminosity and claim that the results from this method are consistent with that of \citet{Liodakis_2018}. \citet{Ye_2023} combined the results from \citet{Liodakis_2018} and \citet{Zhang_2020}, with the Fermi-LAT data for BLL to estimate the Doppler factor of FRI radio galaxies within a unification scenario.

The data from \citet{Zhang_2020}, and \citet{Ye_2023} were combined with \citet{Chen_2018}, and provide the Doppler factor of 15 among the 26 objects found in the Auger Catalog. To get the data for the remaining objects, we use the empiric relation between $f_b \approx \mathcal{D}^{-2}$ and the observed luminosity $L^{iso}$,
\begin{equation}
    \log f_b = (-0.21 \pm 0.03)\log L^{iso} + (7.67 \pm 1.54)
\end{equation}
found by \citet{10.1093/mnras/stad065} using the data from \citet{Liodakis_2018} for FSRQs, BLLs, $\gamma$NLS1s, and radio galaxies. This expression uses $L^{iso}$ calculated from the 4FGL-DR4 Fermi Catalogue, between $0.1-100$~GeV, while we are using the 4FHL catalog to weight the UHECR flux.

\linelabel{fline}

\movetabledown=5.5cm 
\begin{rotatetable} 
\begin{deluxetable}{lccccccccc} 
\tablecaption{Relevant data for the AGNs used in the analysis.\label{tab:AGNs.prop}} 
\tabletypesize{\scriptsize}
\tablehead{ 
 AGN\tablenotemark{a} & Class & (RA, DEC)\tablenotemark{b} & Distance (Mpc) &$F_{3fgl} / 10^{-11}$\tablenotemark{c} & $F_{3fhl}/ 10^{-12}$\tablenotemark{d} & $\mathcal{D}$ & Ref.\tablenotemark{e}& $\mathcal{D}^{-2} F_{3fhl} / 10^{-12} $ & $\mathcal{D}^{-4} F_{3fhl}  / 10^{-12}$} 
 \startdata 
 CenA & RG & (201.37, -43.02) & 3.68 & 5.86 & 7.41 & 1.0 & Y23 & 7.41 & 7.41 \\ 
 M87 & RG & (187.71, 12.39) & 16.7 & 1.64 & 9.55 & 1.3 & Y23 & 5.40 & 3.05 \\ 
 FornaxA & RG & (50.67, -37.21) & 20.4 & 0.61 & 2.59 & 1.0 & Y23 & 2.59 & 2.59 \\ 
 CenB & RG & (206.7, -60.41) & 55.2 & 2.04 & 2.12 & 2.3 & Y23 & 0.40 & 7.59$\times10^{-2}$ \\ 
 NGC1275 & RG & (49.95, 41.51) & 78.0 & 29.13 & 47.97 & 7.5 & Y23 & 0.86 & 1.52$\times10^{-2}$ \\ 
 IC310 & RG & (49.18, 41.32) & 83.2 & 0.33 & 6.90 & 1.9 & Y23 & 1.89 & 0.52 \\ 
 TXS0149+710 & BCU & (28.36, 71.25) & 103.3 & 0.25 & 3.72 & 2.2 & Y23 & 0.76 & 0.16 \\ 
 NGC1218 & RG & (47.11, 4.11) & 124.7 & 0.93 & 2.42 & 3.4 & Y23 & 0.22 & 1.90$\times10^{-2}$ \\ 
 Mkn421 & BLL & (166.1, 38.21) & 133.7 & 40.44 & 437.47 & 14.5 & Z20 & 2.08 & 0.99$\times10^{-2}$ \\ 
 Mkn501 & BLL & (253.47, 39.76) & 152.1 & 11.77 & 156.19 & 5.7 & Z20 & 4.84 & 0.15 \\ 
 TXS0128+554 & BCU & (22.81, 55.75) & 162.9 & 0.61 & 0.99 & 5.1 & C23 & 0.38$\times10^{-1}$ & 0.14$\times10^{-2}$ \\ 
 CGCG050-083 & BCU & (235.89, 4.87) & 178.6 & 0.51 & 3.60 & 5.1 & C23 & 0.14 & 0.52$\times10^{-2}$ \\ 
 1RXSJ022314.6-111741 & BLL & (35.81, -11.29) & 182.8 & 0.26 & 1.24 & 4.8 & C23 & 0.54$\times10^{-1}$ & 0.23$\times10^{-2}$ \\ 
 1ES2344+514 & BLL & (356.76, 51.69) & 196.8 & 3.17 & 30.74 & 2.7 & C18 & 4.22 & 0.57 \\ 
 PMNJ0816-1311 & BLL & (124.11, -13.2) & 200.4 & 1.60 & 16.07 & 5.9 & C23 & 0.46 & 1.30$\times10^{-2}$ \\ 
 Mkn180 & BLL & (174.11, 70.16) & 203.2 & 1.25 & 15.25 & 1.4 & C18 & 7.78 & 3.97 \\ 
 1ES1959+650 & BLL & (299.97, 65.16) & 211.8 & 10.50 & 67.74 & 7.3 & C23 & 1.27 & 2.37$\times10^{-2}$ \\ 
 SBS1646+499 & BLL & (251.9, 49.83) & 212.8 & 1.45 & 1.89 & 5.8 & C18 & 0.56$\times10^{-1}$ & 0.17$\times10^{-2}$ \\ 
 APLibrae & BLL & (229.42, -24.37) & 216.8 & 7.53 & 20.33 & 7.1 & C23 & 0.40$\times10^{-1}$ & 0.80$\times10^{-2}$ \\ 
 TXS0210+515 & BLL & (33.55, 51.77) & 218.8 & 0.54 & 6.21 & 5.4 & C23 & 0.21 & 0.73$\times10^{-2}$ \\ 
 3C371 & BLL & (271.71, 69.82) & 225.9 & 3.29 & 4.29 & 4.0 & Z20 & 0.27 & 1.71$\times10^{-2}$ \\ 
 PKS1349-439 & BLL & (208.24, -44.21) & 228.0 & 0.80 & 0.87 & 5.7 & C23 & 0.27$\times10^{-1}$ & 0.08$\times10^{-2}$ \\ 
 1RXSJ020021.0-410936 & BLL & (30.09, -41.16) & 234.4 & 0.20 & 2.72 & 4.9 & C23 & 0.11 & 0.46$\times10^{-2}$ \\ 
 PKS0625-35 & BLL & (96.78, -35.49) & 238.8 & 1.54 & 12.38 & 6.8 & Y23 & 0.26 & 0.56$\times10^{-2}$ \\ 
 1ES2037+521 & BLL & (309.85, 52.33) & 238.8 & 0.55 & 3.72 & 5.5 & C23 & 0.12 & 0.41$\times10^{-2}$ \\ 
 PKS0521-36 & BLL & (80.76, -36.46) & 241.0 & 5.25 & 5.52 & 7.0 & C23 & 0.11 & 0.23$\times10^{-2}$ \\ 
 \enddata 
 \tablecomments{ 
 \tablenotetext{a}{Associated source found in the Fermi-LAT Catalog.} 
 \tablenotetext{b}{Equatorial coordinates, in degrees} 
 \tablenotetext{c}{Energy Flux from Fermi 3FGL-DR4 catalogue (0.1 - 100 GeV), in erg cm$^{-2}$~s$^{-1}$, \url{https://heasarc.gsfc.nasa.gov/W3Browse/fermi/fermi3fgl.html}} 
 \tablenotetext{d}{Energy Flux from Fermi 3FHL catalog (10 GeV - 1 TeV), in erg cm$^{-2}$~s$^{-1}$, \url{https://heasarc.gsfc.nasa.gov/W3Browse/fermi/fermi3fhl.html}} 
 \tablenotetext{e}{Reference: C18=\citet{Chen_2018}, Z20=\citet{Zhang_2020}, Y23=\citet{Ye_2023}, C23=fit from \citet{10.1093/mnras/stad065}}} 
\end{deluxetable} 
\end{rotatetable}

\end{document}